\documentclass[conference]{IEEEtran}

\usepackage{cite}

\usepackage{multirow}

\ifCLASSINFOpdf
  \usepackage[pdftex]{graphicx}

\else

\fi

\usepackage[shortlabels]{enumitem}

\usepackage{amsmath}

\usepackage{array}
\usepackage{multirow}
\usepackage[table,xcdraw]{xcolor}

\ifCLASSOPTIONcompsoc
 \usepackage[caption=false,font=normalsize,labelfont=sf,textfont=sf]{subfig}
\else
 \usepackage[caption=false,font=footnotesize]{subfig}
\fi

\usepackage{url}

\usepackage{multirow}
\usepackage[normalem]{ulem}
\useunder{\uline}{\ul}{}

\usepackage{algpseudocode}
\usepackage{algorithm}
\usepackage{amsfonts}
\usepackage{multirow}
\usepackage{breqn}
\usepackage{hyperref}
\usepackage{multirow}
\usepackage[table,xcdraw]{xcolor}
\usepackage[table,xcdraw]{xcolor}

\usepackage{subfig}

\begin{document}

\title{Failure Identification from Unstable Log Data using Deep Learning}

\author{\IEEEauthorblockN{Jasmin Bogatinovski\IEEEauthorrefmark{1},\
Sasho Nedelkoski\IEEEauthorrefmark{1},
Li Wu\IEEEauthorrefmark{1},
Jorge Cardoso\IEEEauthorrefmark{2},
Odej Kao\IEEEauthorrefmark{1} 
}
\IEEEauthorblockA{\IEEEauthorrefmark{1}Distributed and Operating Systems, 
Technical University Berlin, Germany}
\IEEEauthorblockA{\IEEEauthorrefmark{2}Huawei Munich Research, Munich, Germany \\ \{jasmin.bogatinovski, odej.kao\}@tu-berlin.de}}

\IEEEoverridecommandlockouts
\IEEEpubid{\begin{minipage}{\textwidth}\ \\[12pt]
		\copyright 2022 IEEE. Permission from IEEE must be obtained for all uses, in any \\
       	 current or future media, including reprinting/republishing this material for \\
       	 advertising or promotional purposes. This paper is accepted at IEEE \\ CCGrid 2022.
       	 For citations use references from the conference proceedings.
	\end{minipage}}

\maketitle

\begin{abstract}
The reliability of cloud platforms is of significant relevance because society increasingly relies on complex software systems running on the cloud. To improve it, cloud providers are automating various maintenance tasks, with failure identification frequently being considered. The precondition for automation is the availability of observability tools, with system logs commonly being used. The focus of this paper is log-based failure identification. This problem is challenging because of the instability of the log data and the incompleteness of the explicit logging failure coverage within the code. To address the two challenges, we present CLog as a method for failure identification. The key idea presented herein based is on our observation that by representing the log data as sequences of subprocesses instead of sequences of log events, the effect of the unstable log data is reduced. CLog introduces a novel subprocess extraction method that uses context-aware neural network and clustering methods to extract meaningful subprocesses. The direct modeling of log event contexts allows the identification of failures with respect to the abrupt context changes, addressing the challenge of insufficient logging failure coverage.  Our experimental results demonstrate that the learned subprocesses representations reduce the instability in the input, allowing CLog to outperform the baselines on the failure identification subproblems -- 1) failure detection by 9-24\% on F1 score and 2) failure type identification by 7\% on the macro averaged F1 score. Further analysis shows the existent negative correlation between the instability in the input event sequences and the detection performance in a model-agnostic manner.
\end{abstract}


\IEEEpeerreviewmaketitle
\section{Introduction}
Cloud systems are a mixture of complex multi-layered software and hardware. They enable applications of ever-increasing heterogeneity and complexity powering different technologies such as the Internet of Things, distributed processing frameworks,  databases, virtual reality, among others. The emergence of complexity within the cloud relates to diverse maintenance challenges, with an important challenge of being prone to failures~\cite{CloudProperties}. The failures have a significant impact on the performance affecting user experience and leading to economic losses~\cite{surveyLogAnalysis}. Therefore, accurate and timely failure identification is crucial for enhancing the reliability of the cloud and its services. 

Cloud providers are considering many approaches to address the problem of failure identification, commonly by adopting various data-driven methods~\cite{He2020SurveyLogMining}. Their fundament resides in the available monitoring data, with log data (logs for short) commonly being utilized. Other monitoring data, like key performance indicator metrics (KPI, e.g., memory utilization, I/O bytes), provide clues for detecting failures, however, they do not provide a verbose description of the type of the failure~\cite{LogClass2021} making the failure identification incomplete. For example, a sharp increase in the curve of memory utilization only indicates that the memory utilization increases, but it cannot tell why it happens in isolation. In comparison, logs are textual data recording events with different granularity, providing human-understandable clues for the failure and its type. For example, from several repetitions of the two consecutive log lines, “$l_1$: Interface changed state to up.” and "$l_2$: Interface changed state to down.", operators can detect a failure in the system, assign its type “Interface Flapping”, conclude that the interface is flapping and obtain clues for potential root causes (\textit{bad cable connection} a common suspect in this example). A single log is composed of a static event template describing the event (e.g., "Interface changed state to $\langle*\rangle$.") and parameters (e.g., up) giving variable event information. 

The focus of this study is the problem of log-based failure identification. Traditionally, it is addressed by manual analysis, like keyword search of failure words (e.g., "fail") or log levels with great severity (e.g., "error")~\cite{LogCluster2016}. Owning to~the unprecedented development of the cloud systems, logs are consistently generated in large volumes (several TB per day~\cite{He2020SurveyLogMining}), making the task of manual log-based diagnosis cumbersome. Thereby, automatic approaches for log-based failure identification are increasingly researched and adopted~\cite{LogRobust, HeShenlin2016Log3C, deepLogAnomalySurvey, UniLog2021, Logsy}.

\begin{figure}[!h]
\centering
\includegraphics[width=0.65\columnwidth]{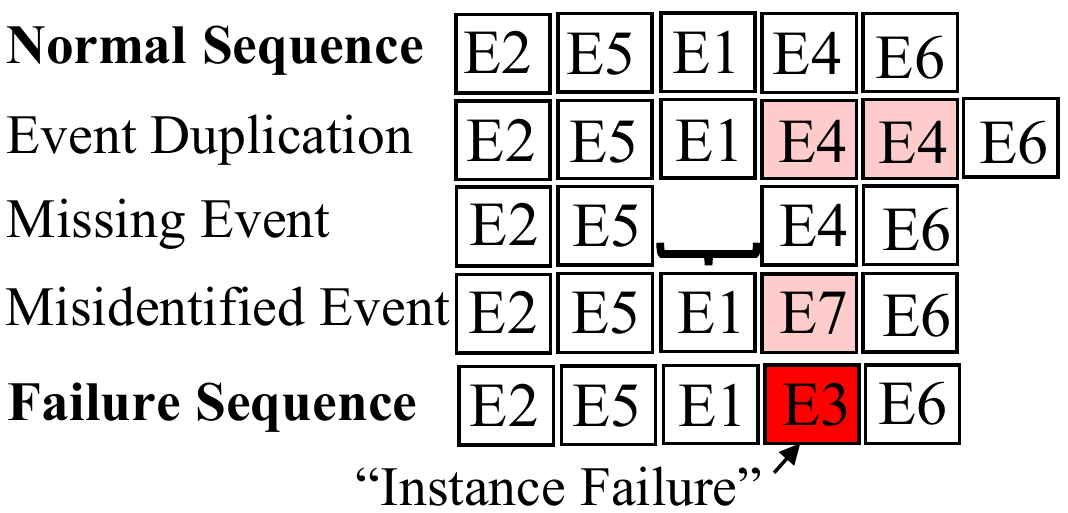}
\caption{Examples of unstable and failure log event sequences.}
\label{fig:challanges}
\end{figure}

Current approaches 1)~identify failures from single log lines~\cite{Logsy} or 2)~exploit groups/contexts of log events (i.e., co-occurring event templates/events) in form of \textit{log sequences}~\cite{DeepLog} (i.e., series of event templates with external identifier) and \textit{count vectors}~\cite{HeShenlin2016Log3C}. Depending on the assumed input, different challenges emerge. When considering groups of log events, the challenge of unstable log sequences occurs~\cite{LogRobust}. Unstable log sequences are sequences from the same type of workload execution having slightly different sequential structures. \figurename~\ref{fig:challanges} depicts examples of unstable sequences caused by different reasons. In the sequences of events denoted with "Event Duplication" and "Missing Event", the original sequence $(E2, E5, E1, E4, E6)$ is modified by repeating a single event "E4" or dropping the event "E1", accordingly. The two sequences still represent normal system behaviour but have slightly different structures. Such problems are common in cloud systems where the log data is analyzed at a central place. The network errors, limited throughput, or storage issues are referenced causes for events repeating or dropping. There are other sources of instabilities (e.g., the preprocessing of raw logs), altering the normal log sequences in a similar way. Notably, the instability causes similar properties of the unstable normal and failure sequences (e.g., shortened lengths or contexts differ in a single event), making it harder to distinguish them from one another. For example, the two sequences in \figurename~\ref{fig:challanges}, labeled "Misidentifying Event" and "Failure", differ just within one event on the fourth position ("E7" and "E3"). The first arises due to an error in the log event preprocessing, while the second is because the template describes a failure event. Therefore, the instability inflicts a modeling challenge and increases the entropy in the data. From modeling perspective, this requires accounting for the unequal importance of the log events within the contexts, impairing the detection performance otherwise~\cite{LogRobust}.

The methods using single log lines do not suffer from the problem of unstable logs, predominantly due to incorporating information about the semantics of the log events~\cite{deepLogAnomalySurvey}. These approaches demonstrate strong performance~\cite{Logsy} however, they cannot detect failures that are not explicitly logged. For example, in the aforenamed failure with type "Interface Flapping" (with logs $l_1, l_2$), none of the two logs has a log level with greater severity (i.e., "error" or "critical"), nor do they explicitly describe a failure. The failure can be detected just within the context of several repetitions of the specific pair of logs. Furthermore, these types of contextual failures occur often. For example, for the release Pike (version 3.12.1) of a popular cloud resource managing system OpenStack, there are more than 20\% of failures not explicitly logged within a single log line~\cite{cotroneo2019bad}. Acknowledging that developers have an insufficient understating of the complexities of the running system environment during development results in insufficient failure logging coverage~\cite{He2020SurveyLogMining}. Conclusively, the failures that do not manifest in individual lines make the failure identification possible only within the context of other logs (e.g., presence/absence of frequently co-occurring logs).

\textbf{Contributions.}  \textbf{1)} To overcome the two challenges, in this paper as the \textit{main contribution}, we introduce CLog -- a method for log-based failure identification. The key idea of CLog is reducing the instability of the input log event sequences by representing them as sequences of subprocesses (i.e., groups of similar contexts). Since subprocesses represent contexts (co-occurring log event templates), their number is significantly smaller than the event number used to represent the sequences. The two key benefits of the change in the representation are that -- a) by representing the log event sequences by a smaller number of subprocesses, we directly reduce the entropy in the input representation, reducing the effect of the unstable log sequences; b) the modeling of contexts allows detecting failures in terms of abrupt context changes, addressing the challenge of insufficient logging failure coverage. The challenge that arises is the extraction of subprocesses. \textbf{2)}~To address it, we contribute a novel method for unsupervised subprocesses extraction based on context-aware deep learning and clustering methods. \textbf{3)} Our experimental results on two datasets from OpenStack (with 172 failures) demonstrate that CLog outperforms the baselines on the two failure identification subproblems: failure detection (by 9-24\% on F1 score) and failure type identification (by 7\% on macro average F1 score). By injecting unstable event sequences, we show CLog's robust performance dropping by just 6\% under a severe ratio of unstable sequences. \textbf{4)} Finally, we contribute by open-sourcing the datasets and method for reproducibility purposes and fostering the research on this practically relevant problem.

The remaining of the paper is structured as follows. Section~\ref{Prelim} gives the key observation for the approach, alongside the problem definition. Section~\ref{method} describes the proposed methodology. Section~\ref{exp} discusses the experimental results in response to four research questions. Section~\ref{rw} discusses the related work for the two sub-problems of failure detection and failure type identification. Section~\ref{conc} concludes the paper and gives directions for future work.

\section{Preliminary}\label{Prelim}

\subsection{Problem Definition}
In this paper, we address the problem of log-based failure identification~\cite{LogCluster2016}. We decompose the problem into two subproblems, i.e., (1) failure detection and (2) failure type identification, defined in the following.

\textbf{Failure Detection (FD)}. Let $L=\{l_1, l_2 \dots l_i \dots l_n\}$ be a set of $n$ time-ordered logs from cloud services, and there exist an index set $\mathbb{J} \in \mathbb{N}$ capturing dependency relation between the logs, i.e., $s_j = (l_{ji} \in L | j \in \mathbb{J})$, where  $l_{ji}$ denotes individual log of the sequence $s_j$. Further, we assume that there exist a function $p^{+}$ denoting the normality score of the sequence $p^{+}(\phi(s_j)): \mathbb{R}^d\mapsto\mathbb{R}$, where $\phi: \mathbb{S} \mapsto \mathbb{R}^d$ is the representation function of sequence $s_j$ into $d$-dimensional numerical vector space, and $\mathbb{S}$ is the available sequence set. The task of failure detection is defined as finding the set $A=\{s_j \in \mathbb{S} |a_1<p^{+}(s_j) ||p^{+}(s_j)>a_2, j \in \mathbb{J}\}$, where $a_1, a_2$ are constants such that $a_1<a_2$. Although the individual logs $l_i$ in the sequence $s_j$ can describe normal events, the overall sequence can denote a failure. The index set $\mathbb{J}$ in the context of logs can represent task ID, process ID, or workload ID. It can be given apriori (as considered here) or reconstructed by an additional procedure. We assume that the majority of the log messages $l_i$ and the sequences $s_j$ describe normal system behaviour. 

\textbf{Failure Type Identification (FTI)}. Given a set of detected failure sequences $A$ and the set of failure types identifiers $T=\{t_1, t_2 \dots t_{w}\}$, where $w$ denotes the number of unique failure type identifiers, the task of failure type identification is finding a function $f(\phi(s_i)): A \mapsto T$.

\textit{\textbf{Failure Identification (FI)}.} Given the sets $L, T, \mathbb{S}$, and $\mathbb{J}$, the task of failure identification is finding the set $\tilde{A}=\{(s_1, t_1), \dots (s_i, t_i) \dots (s_{|A|}, t_{|A|})\}$, where the failure sequence $s_i$ is detected by estimating the normality score function $\tilde{p}^{+}(\tilde{\phi}(s_i))$ and the thresholds $\tilde{a_1}$ and $\tilde{a_2}$, while its type $t_i$ is identified by the estimate of $\tilde{f}(\tilde{\phi}(s_i))$. The estimates of $\tilde{\phi}(s_i)$, $\tilde{p}^{+}(\tilde{\phi}(s_i))$ and $\tilde{f}(\tilde{\phi}(s_i))$ further are used for representing, and failure identification on novel sequences. CLog addresses the problem by finding suitable representation for the sequences $\tilde{\phi}(s_i)$ (Section~\ref{PLog}), which are used to find estimates for $\tilde{p}^{+}$ (Section~\ref{FD}) and $\tilde{f}$ (Section~\ref{FTI}).

\begin{figure}[!h]
\centering
\includegraphics[width=0.75\columnwidth]{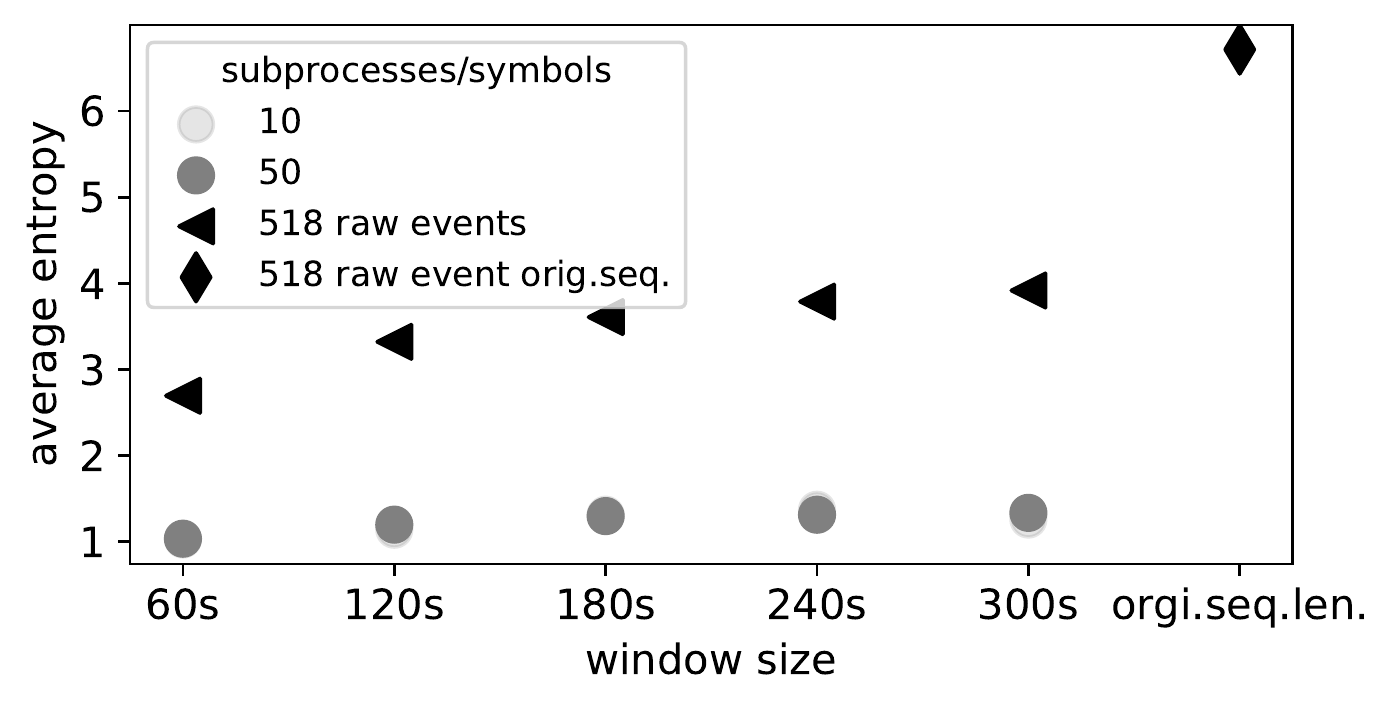}
\caption{Entropy (measuring instability) increases with more unique symbols.}
\label{fig:motivation}
\end{figure}
\subsection{Key Observation}~\label{keyObservation}
The key observation CLog relies on is reducing the entropy of the input representation of log sequences by representing them as sequences of subprocesses instead of log events.  For example, a sequence of log events  $s_i = (E_1, E_2, E_5, E_3, E_1)$ with a task ID $i$, where each of $\{E_1, E_2, E_3, E_5\}$ denotes log event template, can equivalently be represented as sequence of two subprocesses/symbols $s_i=(S_a, S_b)$ such that $S_a=(E_1, E_2, E_5)$, and $S_b=(E_3, E_1)$, while $S_a$ and $S_b$ are referred to as subprocesses. \figurename~\ref{fig:motivation} depicts the impact of changing the log event sequence representation into sequences of subprocesses on data from OpenStack. It is seen that with increasing the grouping window size, the number of log events in the subprocesses increases, and the average entropy over all the sequences in the data also increases (the triangles). The entropy is the highest when the whole sequence is represented with individual log events (the diamond). Notably, when the sequences are represented with subprocesses, the number of symbols used to represent the sequences is smaller (the circles), and the average entropy is reduced. It implies a reduction of the sensitiveness over the individual logs, effectively reducing the effect of the instability in the log event sequences. An important goal of CLog is to learn subprocesses by preserving their characteristics, i.e., by learning context-aware event sequence groups.

\section{CLog: Method for Log-based Failure Identification}\label{method}
\begin{figure}[!h]
\centering
\includegraphics[width=1.0\columnwidth]{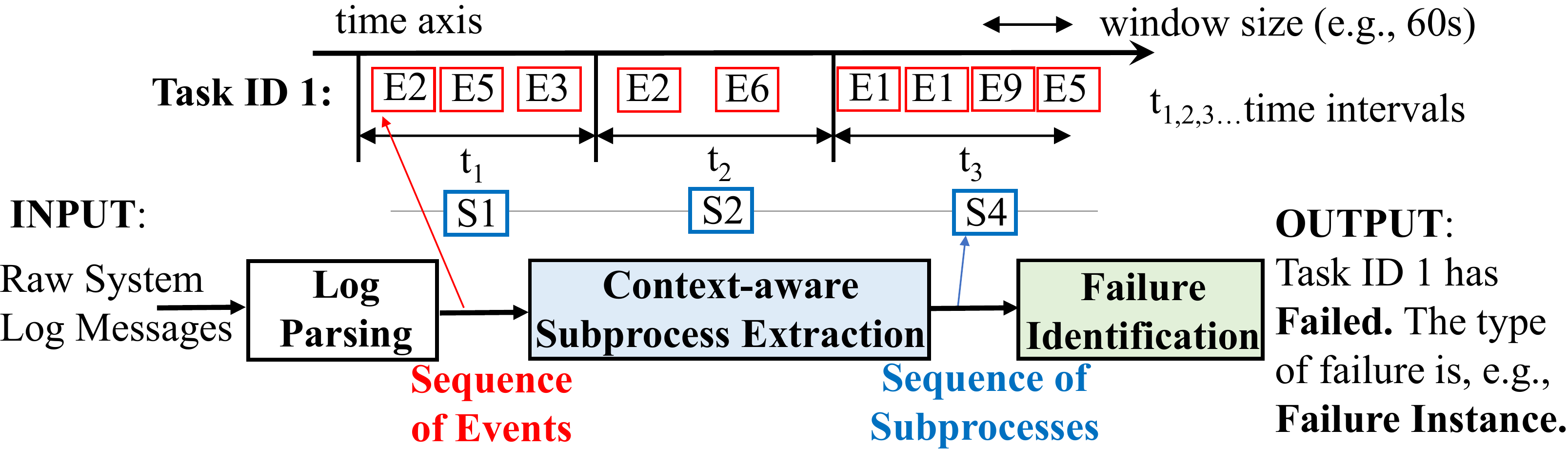}
\caption{CLog overview.}
\label{fig:overview}
\end{figure}

To address the problem of failure identification, we propose CLog. \figurename~\ref{fig:overview} gives an overview of the method. It has three parts 1) log parsing, 2) context-aware subprocesses extraction, and 3) failure identification. The log parsing, as a general preprocessing procedure in log analysis~\cite{ParsingSurvey}, extracts the event templates from the incoming raw log event sequences, transforming the raw log sequences into sequences of log events. The event sequences are processed by the context-aware subprocesses extraction part, converting them into sequences of subprocesses. This part leverages our observation that by representing the log sequences on a level of subprocesses, the entropy of the representation sequence is smaller. This increases the stability of the input and reduces the impact of the unstable logs. Finally, the processed log sequences of subprocesses are given as input into the failure identification part. The latter is composed of two modules (a) failure detector and (b) failure type identification (FTI). The goal of the failure detector is to detect the failure sequences of subprocesses. The FTI module further identifies the failure types based on operators experience. CLog has two modes of operation: offline and online. During the offline phase, the parameters of the log parser, context-aware subprocesses part, and failure identification parts are learned, and the models are induced and stored. In the online phase, the stored models are loaded and used to identify failures. In the following, we describe the internal mechanisms of the three parts of CLog in detail.
\subsection{Log Parsing}
The generated raw system logs are unstructured. Since we are interested in modeling the sequences of events, as the first step, we extract the event templates from the raw logs by applying automatic log parsing. Log parsing decouples the log templates from log parameters (the variable part in a log), directly extracting input in useable modeling format. We prefered automatic log parsing because the alternatives (e.g., regular expression and "grock" patterns), although successful for parsing the templates, are system-specific, requiring frequent updates, which makes them challenging for maintenance~\cite{ParsingSurvey}.

While there are many log parsers available, CLog utilizes a tree-based parser Drain~\cite{Drain}. Zhu et al.~\cite{ParsingSurvey} identified Drain among the most efficient in comparison to 12 other parsers on ten benchmark datasets from diverse software systems. The three main properties making it popular and widely adopted are its correctness, efficiency and intuitive meaning of the hyperparameters (making the tuning process undemanding). Once parsed, the events are organized in sequences by CLog's hyperparameter \textit{window size}, corresponding to the arriving time interval and task ID of the events, and are proceeded as output towards the context-aware subprocess extraction part. 

\subsection{Context-aware Subprocess Extraction}~\label{PLog}
\begin{figure}[!h]
\centering
\includegraphics[width=1\columnwidth]{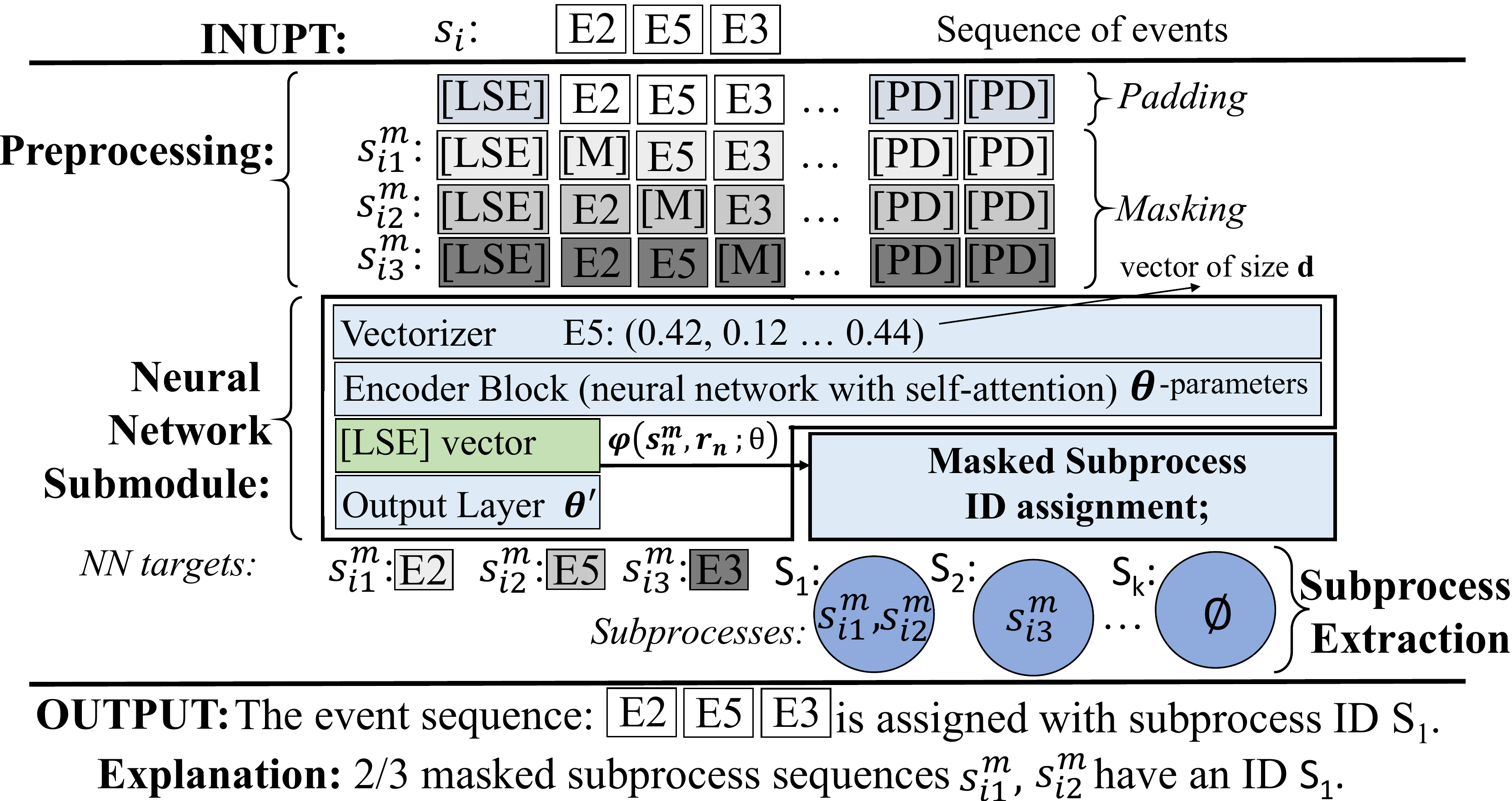}
\caption{The internal design of context-aware subprocess extraction part (with a detailed explanation of a running example).}
\label{fig:encoderBlock}
\end{figure}

The context-aware subprocess extraction is the central part of the method. Its goal is the extraction of subprocesses from the parsed log event sequences. By representing an execution workload (with a task ID) on a higher-level granularity, i.e., by subprocesses, we reduce the entropy in the input, addressing the problem of unstable log event sequences. This part combines context-aware neural network and clustering methods to learn explicit relationships between the events within the event sequences preserving their local properties. \figurename~\ref{fig:encoderBlock} depicts the overall design of the context-aware subprocess extraction part with a running example. Conceptually, it is composed of three submodules -- (1) preprocessing submodule that transforms the input sequences into a format suitable for learning, (2) neural network learning module which is combined with a batched kmeans method to learn subprocesses in an unsupervised manner, (3) subprocess extraction module that assigns a unique subprocess identifier to the input event sequence. In the following, we describe the submodules. 

\subsubsection{Preprocessing Submodule} The goal of the preprocessing submodule is to preprocess the input log event sequences in a unified format suitable for the neural network. It has two components: padding and masking. 

\textbf{Padding}. The padding component receives the sequences of log events as input, with each event represented by a unique identifier (e.g., integer). We refer to it as a token. The sequences of events in a given time interval are different in length. However, the neural network requires a fixed-size representation of the input. The padding component specifies a hyperparameter $max\_length$ and appends each of the shorter log sequences with a dedicated token $[PD]$ up to $max\_length$ to enforce fixed-size representation. The longer sequences (having more than $max\_length$ events) are truncated. Notably, we add a dedicated token $[LSE]$ (Log Sequence Embedding) at the beginning of each sequence. During learning, we enforce the sequence token representations to propagate through the upper layers in the network via the $[LSE]$ token. Thereby, $[LSE]$ attends over all the tokens from the sequence and summarizes the relevant context during learning. The $[LSE]$ token serves as a sequence vector representation later used to group contexts and identify subprocesses. The output of the padding module is the prepended and padded event sequence. 

\textbf{Masking}. To learn context-aware groups, we consider a general self-supervised learning task from natural language processing (NLP) research called Masked Language Modeling (MLM)~\cite{BERT}. To apply the MLM task, the masking component is processing the prepended and padded log event sequences in a suitable format. More specifically, as input, it receives the prepended and padded log event sequences and outputs a set of pairs of masked log event sequences and original masked events. Masked log event sequences are sequences of log events created by replacing all of the events from an original log sequence with a special $[M]$ (masked) token. For example, for the input sequence $(E_2, E_5, E_3)$, one masked sequence is $(E_2, [M], E_3)$, with $E_5$ being the original masked event. There are three masked event sequences for this example. $[LSE]$ and $[PD]$ tokens are not affected by the masking procedure. During training, a masked sequence is given as input to the neural network, while the original masked token is used as the prediction target. By predicting the mask from the co-occurring tokens, the method learns the most important events from the surrounding context, extracting context-aware representations. Note that by this procedure single input sequence is multiplied several times. We keep track of the origin (the input event sequence) of each masked sequence and use it to extract its corresponding subprocess identifier (see Section~\ref{SE}). 

\subsubsection{Neural Network Submodule} The neural network submodule learns context-aware groups of masked log sequences. It implements a neural network following the design of a self-attention encoder of the Transformer~\cite{BERT} architecture. The advantage given by this architectural choice resides in its capability to learn the contextual information between the input events. When learning the parameters of the network, guided by a carefully designed cost function, the model learns local relationships between the events based on their co-occurrence extracting useful contextual features. The neural network submodule has four components: vectorizer, encoder block, output layers and masked subprocess id assignment. The \textbf{vectorizer} transforms the masked input event sequences of tokens into numerical vector sequences. These vectors are called \textit{event embeddings} and are part of the training procedure. At the beginning of the training procedure, the event vectors are randomly initialized and are updated during training. This way, they learn contextual information about the events.

The \textbf{encoder block} is composed of a self-attention encoder layer. The self-attention extracts co-occurring information by weighting the input vector embeddings by their similarity to all the other embeddings in the given context. Combining the self-attention with the MLM learning task modifies the parameters of the network to learn the context of the original masked event and extract sequential properties. The hyperparameters of the encoder are the model size (denoted by $d$), the number of encoder layers, the number of heads, and the dropout ratio (used to prevent overfitting). We reference the reader to Devlin et al.~\cite{BERT} for the specific details on self-attention neural networks. Particularly interesting is the embedding of the $[LSE]$ token. Since $[LSE]$ serves as an embedding of the sequence, it learns contextual properties of the masked input sequences. The output from the encoder is the vector embedding of the \texttt{$[LSE]$} token for each of the masked sequences, proceeded towards the output layer.

The \textbf{output layer} is composed of two layers with nonlinear activation (RELU is used). The purpose of it is to map the masked sequence embedding vector \texttt{$[LSE]$} of size $d$, to a vector with a size corresponding to the total number of events/tokens $C$. The output of this layer is used to calculate the loss. As an optimization loss function, we use categorical cross-entropy. Notably, during the execution of a workload, some events occur only once (e.g., "notification of successful creation of a VM") while others in greater frequency (e.g., HTTP or RPC calls). When using the original loss formulation on the MLM task, the less frequent events will be averaged out, resulting in missing important information. To account for the imbalances of the distribution of the events, we use weighted categorical cross-entropy given in Eq.~\ref{weightedCrossEntropy} as follows:
\begin{dmath}
    J_m(\psi(\mathbf{s^{m}_{n,c}};\theta,\theta^{'}), y^{m}_{n,c}; \mathbf{w}) = \frac{1}{|C|} \sum_{c=1}^{C}-w_{c}y^{m}_{n,c}  log\frac{exp(\psi(\mathbf{s^{m}_{n,c}};\theta, \theta^{'}))}{\sum_{i=1}^{C}exp(\psi(\mathbf{s^{m}_{n, i}};\theta, \theta^{'}))}
    \label{weightedCrossEntropy}
\end{dmath}
where $\psi$ denotes the function modeled by the neural network, $\theta$ and $\theta{'}$ are the parameters of the encoder, and the output layer accordingly, $\mathbf{s^{m}_{n,c}}$ is a masked sequence obtained from the $n$-th input sequence $\mathbf{s_n}$, $y^{m}_{n,c}$ is the original masked event/token, $C$ denotes the total token numbers and $w_c$ represents the weight of an individual token. The weights ($\mathbf{w}$ -- a weights vector) are assigned such that the less frequent events have weight values closer to 1, as opposed to the frequent ones that have values closer to 0. Therefore, we optimize for preserving the correct predictions on the infrequent events, addressing the challenge of the imbalance of the event frequency distribution. 

\textbf{Masked Subprocess ID assignment}. The masked subprocess ID assignment receives the vector embedding of the \texttt{$[LSE]$} token as input. 
It applies the mini batched kmeans algorithm~\cite{TowardsKmeansFriendlyClustering} to group the embeddings of the masked event sequences into a predetermined number of $\mathbf{k}$ subprocesses/centroids identifiers. The mini batched kmeans algorithm is a commonly used method for identifying similar instance groups in an unsupervised way. While the goal of the encoder block is to learn context-aware representations, the mini batched kmeans complements it by extracting similar context groups, enabling the extraction of subprocesses. We used mini batch kmeans because it allows per batch update of the network $(\theta$ and $\theta^{'})$ and clustering parameters (\textbf{M}) as opposed to the classical kmeans method. To group the contexts, kmeans optimizes the loss given in Eq.~\ref{kmeansLoss} by altering between two steps: 1) updating a centroid $m_{k}$ as the average of the embeddings currently assigned to it, and 2) reassignment of the embeddings to the nearest newly calculated centroid. 
\begin{equation}
    J_k(\phi(\mathbf{s^{m}_n}, \mathbf{r_n};\theta),\mathbf{M}) = ||\phi(\mathbf{s^{m}_n};\theta) - \mathbf{r_n}\mathbf{M}||_2
    \label{kmeansLoss}
\end{equation}
where $\mathbf{M}\in\mathbb{R}^{kxd}$ represent the matrix of subprocess context-group (interchangeably referred to as centroids), while $\mathbf{r_n}$ is an indicator vector of discrete values (0's and 1's) with just one element set to one, corresponding to the membership of the masked sequence $\mathbf{s^{m}_n}$ to a certain centroid $m_{k}$. The number of subprocesses identifiers $\mathbf{k}$ is a hyperparameter.

Finally, we add the two optimization losses as $J = J_m + \lambda J_k$ to obtain the final loss subject to optimization. By combined optimization of the two losses, the parameters of the context-aware subprocess extraction learn local contexts and local-context groups based on their similarity. The role of the hyperparameter $\lambda$ is to ensure learning of correct contexts and correct context-embedding groups by trading off the impact of the two losses. We further discuss the optimization procedure.

\textbf{Optimization}. The optimization is done in two phases: 1) pretraining and 2) joint training. We first describe the \textit{pretraining phase}. Since at the beginning everything is initialized at random, we pre-train the neural network parameters ($\theta$ and $\theta^{'}$) by the weighted cross-entropy loss (Eq.~\ref{weightedCrossEntropy}). That way, the model learns good initial parameters for the encoder while extracting context-aware features for the masked sequences. The pretraining is terminated after observing a lack of improvement in the loss on five consecutive epochs.  At the end of the pretraining, the $[LSE]$ vectors are valid representations of the masked input sequences. Afterwards, the subprocesses prototypes ($\mathbf{M}$) are initialized by kmeans using $[LSE]$ masked sequence embeddings of the training data.

\textit{Joint training (phase 2)}. The joint optimization function has a discrete variable ($\mathbf{r_n}$), making the parameter updates non-trivial. To address this issue, we calculate the gradients by alternating stochastic gradient descent (ASGD)~\cite{TowardsKmeansFriendlyClustering}. ASGD alters the updates of the network parameters and centroids such that, when the network parameters are updated, the centroids are fixed and vice versa. Therefore, the optimization problem does not depend on the discrete variable, enabling the parameter updates. The training of the network parameters and the centroids is done in batches. Eq.~\ref{updateCentroids} is used for centroids update. At each batch, the centroids with newly assigned embeddings are slightly updated based on their distance to the newly calculated centroids.~Additionally, some of the centroids are updated more frequently than others making the loss convergence slower. Inspired by Yang et al.~\cite{TowardsKmeansFriendlyClustering}, we resolve this issue by penalizing the updates with the term~$\frac{1}{c_k}$. $C_k$ counts the number of times a cluster is assigned an embedding during an epoch. The larger the number of assigned embeddings, the smaller is the centroid updated and vice versa. It normalizes the intensity of the centroid update as a learning rate, different for each cluster.
\begin{equation}
m_k \leftarrow m_k - \frac{1}{c_k}(\phi(\mathbf{s^{m}_n};\theta) - m_k)\mathbf{r_n}
    \label{updateCentroids}
\end{equation}

\subsubsection{Subbprocess Exctraction}~\label{SE} The extraction of a subprocess identifier (ID) is done as follows. Given an original input event sequence and the subprocess ID assignments of its masked subsequences, we count the number of occurrences of the subprocesses IDs and divide the counts by the length of the original input event sequence. The subprocess ID with the highest score value is assigned as a subprocess ID for the input event sequence. Intuitively, if the majority of the masked subsequences are assigned with a single subprocess ID, the subprocess ID with the maximal score value is the most relevant for the input event sequence.~\figurename~\ref{fig:encoderBlock} depicts an example of extracting the subprocess $S_1$ for the sequence $(E_2, E_5, E_3)$. 

\subsection{Failure Identification}
\begin{figure}[!t]
\centering
\includegraphics[width=\columnwidth]{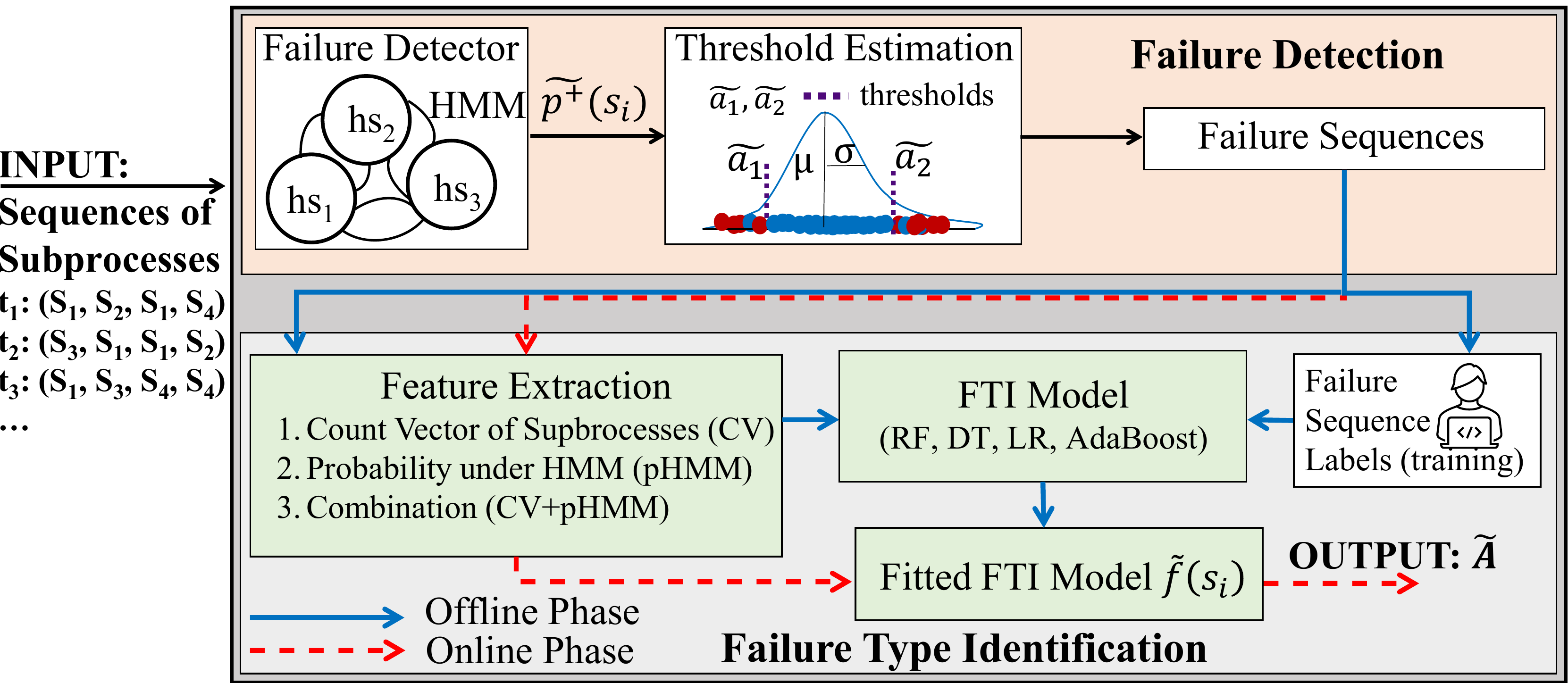}
\caption{Internal architectural design of failure identification part.}
\label{fig:FI}
\end{figure}

The failure identification part is given sequences of subprocesses with the same task ID as input. \figurename~\ref{fig:FI} depicts the internal design. It is composed of two subparts 1) failure detector and 2) failure type identifier. The failure detector detects if the input sequence of subprocesses represents failure. When failure is detected, the sequence proceeds towards the failure type identification part, which identifies the type of failure based on prior historical information. We describe the details in the following. 

\subsubsection{Failure Detection}~\label{FD} As a modeling choice for the \textbf{failure detector}, we considered Hidden Markov Model (HMM)~\cite{KDD2005}. HMM, models the sequences of subprocesses by assuming that the appearance of the next subprocess within the sequence depends only on the current subprocess. The main advantages of HMM are that it directly handles sequential data, does not require further preprocessing of the input, and is fast for both learning and inference (with a reasonably high number of hidden states). To produce normality score estimates for a sequence $\tilde{p}^{+}(\mathbf{s_i})$, we used HMM probability scores ($t(\mathbf{s})$), calculated by marginalizing the probabilities over all the subprocesses of the sequence and the hidden states of the fitted HMM $t(\mathbf{s})=-\log \sum_h q(h)q(\mathbf{\mathbf{s}}|h)$, where $h$ denotes the hidden states, and $q(\textbf{s}|h)$ denotes the likelihood of the subprocess given the hidden state. The normality score estimates for a single sequence $\mathbf{s_i}$ is given in Eq.~\ref{markovModel}, as follows:
\begin{equation}
    \tilde{p}^{+}(\mathbf{s_i}) = (\frac{1}{|\mathbb{V}|}\sum_{s_j}^{|\mathbb{V}|}t(\mathbf{s_j})-t(\mathbf{s_i}))^2
    \label{markovModel}
\end{equation}

where $\mathbb{V}$ is the set of validation normal sequences of subprocesses. The normality score estimate $\tilde{p}^{+}(\mathbf{s_i})$ is a symmetric positive function, given as the spread of the probability of the sequence $\mathbf{s_i}$ under the HMM ($t(s_i)$) from the mean score estimates of the validation data. The parameters of the HMM are learned on the normal training data, thereby, the failure detector models the normal system state. We assume that the normal data is always obtainable from the periods of system operation when there are no issue reports or log events with "error" or "critical" log levels. 
Any sequence with significantly different values for the normality score estimate is \textbf{detected} as a failure. Using the symmetrical property of the normality function, we \textbf{estimate the thresholds} as $\tilde{a}_{1/2}=\mu\pm3\sigma$, where $\mu$ and $\sigma$ are the mean value and standard deviation of the validation score estimates calculated by standard formulas. Thereby, the failure detector is fully unsupervised. The number of hidden states is one hyperparameter of HMM.

\subsubsection{Failure Type Identification}~\label{FTI} Once the failure is detected, the failure sequence proceeds towards the failure type identification module. This module leverages the redundancy property of failures in cloud systems~\cite{LogCluster2016}. This property emerges for various reasons, including temporary failure fixes by developers without addressing the root cause, environmental issues (e.g., machine failure or network disconnections), or running the same system in different environments. Notably, the redundancy implies repetitive patterns in logs, allowing usage of operational information to identify the failure type. The failure type identification subpart has two components (1)~feature extraction and (2) failure type identification model.

The \textbf{feature extraction} processes the sequences of subprocesses in a format suitable for the FTI learning method. Each sequence is represented by a count vector that counts the number of occurrences of a subprocess within the sequence. For example, for the sequence of subprocesses $\mathbf{s}=(S_1, S_3, S_1)$ and total of four subprocesses ($S_1, S_2, S_3, S_4$), the count vector is given as $CV(\mathbf{s})=(2, 0, 1, 0)$. The absence/presence of certain subprocesses from the sequence (e.g., lack of the subprocess with the event "Failure to spawn an instance.") are distinctive features that discriminate among failure types. Therefore, the count vector is a suitable sequence representation. We also considered the normality score estimates from the failure detector as an additional feature (pHMM).

The extracted features are used to fit the \textbf{FTI model} given by $\tilde{f}(s_i)$. FTI learns a multiclass classification model that classifies the input sequences into several predefined types of failures. As an adequate methods we considered several popular multiclass classification methods, i.e., Random Forest (RF)~\cite{Breiman2001}, Decision Tree (DT)~\cite{QuinLan1986}, Logistic Regression (LR)~\cite{LogRegression} and AdaBoost~\cite{Schapire1996}. They show good performance and do not require extensive hyperparameter optimization~\cite{Breiman2001}. The final output of CLog is given as $\tilde{A}=\{(\mathbf{s_j}, t_i )| \mathbf{s_j}\in \mathbb{S}, t_i\in \mathbb{T}, \tilde{p}^{+}(\mathbf{s_j})<\tilde{a_1}|| \tilde{p}^{+}(\mathbf{s_j})>\tilde{a_2}, t_i=\tilde{f}(\mathbf{s_j}), j \in J\}$.

\begin{table*}[!h]
\caption{Dataset Statistics}
\label{tab:properties}
\centering
\resizebox{2\columnwidth}{!}{%
\begin{tabular}{|l|c|c|c|c|c|c|c|}
\hline
Dataset name   & Number of Tasks & No Failure & Assertion Failures & Number of Log Messages & Number of Unique Events & Number of Failure Events & \begin{tabular}[c]{@{}c@{}}Average number of events\\ fault-free task\end{tabular} \\ \hline
OpenStack  & 878             & 706        & 172               & 217534                 & 518                     & 167                      & 1323                                                                               \\ \hline
Syntetic  & 500             & 421-476        & 24-79               & 167215                & 474                     & 123                      & 1309                                                                               \\ \hline
\end{tabular}%
}
\end{table*}
\section{Experimental Evaluation}\label{exp}
In this section, we describe the experimental evaluation. We give details about the experimental design, present and discuss the experimental results in response to four research questions. 

\subsection{Experimental Design}

\subsubsection{\textbf{OpenStack dataset}} To evaluate CLog, we considered a large scale study of failures in OpenStack, introduced in Cotroneo et al.~\cite{cotroneo2019bad}. To the best of our knowledge, it is the most comprehensive publicly available dataset of log failure data from a cloud system. Its strength is the wide range of covered failures following the most common problem reports in the \href{https://bugs.launchpad.net/}{OpenStack bug repository}. The faults are generated by software fault-injection procedure, i.e., modifying the source code of OpenStack and running a predefined workload under fault-injected and fault-free (normal) conditions. 

The considered fault types are grouped into four groups as of following: 1) \textit{throw exception} (method raises an exception in accordance to a predefined API list), 2) \textit{wrong return value} (method returns an incorrect value, e.g., return null reference), 3) \textit{wrong parameter value} (calling a method with an incorrect value for a parameter), and 4) \textit{delay} (method returns the result after a long delay, e.g., caused by hardware failure -- leading to triggering timeout mechanisms or stall). 
As a running workload with a unique task ID, the authors considered the creation of a new instance deployment. This workload configures a new virtual infrastructure from scratch -- it creates VM instances, volumes, key pairs, and security groups, virtual network, assigns instance floating IPs, reboots the instances, attaches the instances to volumes and deletes all resources. Importantly, this comprehensive workload invokes the three key services of OpenStack Nova, Cinder, and Neutron, causing diverse manifestations of the faults as failures.

To generate ground truth labels for the failure state, assertion and API checks are performed at the end of the workload runs. There are three failure types: 1) failure instance, 2) failure SSH and 3) failure attaching volume. While the authors provide information on a granularity of a workload with a task ID, we further labeled the individual logs. More specifically, two human annotators labeled more than 200000 logs to find the ones related to the logged failure. The agreement between the annotators is 0.67 Cohen's Kappa score. \tablename~\ref{tab:properties} gives the detailed statistics of the used data. 

\textit{\textbf{Syntetic dataset:}} To evaluate the robustness of our method in dealing with unstable log data, we have created a synthetic dataset. The data is created similarly as in~\cite{LogRobust}. We start with the normal OpenStack dataset and apply the following three operations to extract failure sequences, i.e., 1) random removal of log events, 2) repetition of a randomly selected log event in the sampled log sequence, and 3) random shuffling the order of several events. To inject unstable log event sequences, we randomly sample 500 log sequences (normal and failed), and in $b$-percentage in the sampled data, we inject the aforenamed operations in random order. 

\subsubsection{Baselines} We compare the failure detection method against three unsupervised baselines (two sequential-based DeepLog~\cite{DeepLog}, HMM~\cite{KDD2005}, and one count-based PCA~\cite{PCA}), and two methods commonly used in practice by developers~\cite{LogCluster2016}. Those are "Log Level", which uses the severity level of the log (i.e., failure exist if the log level is one of "error", "fatal", or "critical") and "Semantic" based on the semantics of a log (i.e., a human identifies the failure as logged in a single log line)~\cite{Hassan2017}. Recent study~\cite{deepLogAnomalySurvey} identifies DeepLog as having a state-of-the-art performance among unsupervised methods. Additionally, we considered a supervised automatic failure identification method LogRobust~\cite{LogRobust}, that requires labels for the severity level of the sequences. For failure type identification, we compare against LogClass~\cite{LogClass2021}, which trains a multiclass model on individual logs to identify failures types. We used task ID failure type alongside the annotations of the single logs to construct a target label and apply this method. 

\subsubsection{Experimental Setup} We conducted the experiments as follows. The hyperparameters of the log parser Drain, i.e., the similarity threshold and depth, were set to 0.45 and 5 as commonly used values for OpenStack logs~\cite{ParsingSurvey}. For \textit{phase 1} the training was performed for a maximal of 200 epochs, and \textit{phase 2} training for a maximal of 20 epochs. As an optimizer, we used SGD with a learning rate set to 0.0001. For the encoder, the model size $d$ was set to 128, with two encoder layers and four heads. To prevent overfitting, we set the dropout rate to 0.01. Experimentally, we find that $\lambda$ with value 0.1 leads to robust results. The optimized hyperparameters of CLog (performed on a separate validation set) are the number of extracted subprocesses, the window size, and the number of hidden states of the HMM. They were selected from the range values of the sets $\{10, 20, 30, 40, 50\}$, $\{60s, 120s, 180s, 240s, 300s\}$ and $\{2, 4, 8, 16\}$ accordingly. The $max\_length$ was set to 32. The hyperparameters of the considered FTI methods set are to their implementation defaults from the \href{https://scikit-learn.org/stable/whats\_new/v0.23.html}{sckit-learn library}. The baselines for failure detection were trained following a survey~\cite{surveyLogAnalysis} of log-based failure detection from software systems. LogClass was trained as in the original paper~\cite{LogClass2021}. The failure detection performance was evaluated on F1, precision and recall as common evaluation metrics, with the failure being a positive label. The same performance scores were used for FTI, with macro averaging over the three failure types. The experiments were conducted on a Linux server with Intel Xeon(R) 2.40GHz CPU and RTX~2080~GPU running with Python~3.6 and PyTorch~1.5.0.

\subsection{Research Questions}
\begin{table*}[!t]
\caption{Comparison of CLog against baselines on failure detection.}
\label{tab:FD}
\centering
\begin{tabular}{c|cccl|ll|l}
\hline
\multicolumn{1}{l|}{Scores/Category} & \multicolumn{4}{c|}{Unsupervised}                                                                                                          & \multicolumn{2}{c|}{Developer Practicies} & \multicolumn{1}{c}{Supervised} \\ \hline
Methods:                              & \multicolumn{1}{c|}{CLog}                     & \multicolumn{1}{c|}{HMM}       & \multicolumn{1}{c|}{PCA}       & DeepLog                  & \multicolumn{1}{l|}{Semantic} & Log Level & LogRobust                       \\ \hline
F1                                    & \multicolumn{1}{c|}{{\ul \textbf{0.94±0.02}}} & \multicolumn{1}{c|}{0.82±0.09} & \multicolumn{1}{c|}{0.77±0.05} & 0.85±0.03                & \multicolumn{1}{l|}{0.81±0.0}  & 0.70±0.02 & 0.96±0.01                       \\
Precision                             & \multicolumn{1}{c|}{{\ul \textbf{0.97±0.03}}} & \multicolumn{1}{c|}{0.8±0.11} & \multicolumn{1}{c|}{0.82±0.07} & 0.78±0.02                & \multicolumn{1}{l|}{{1.0±0.0}}  & 0.74±0.02 & 0.94±0.02                       \\ 
Recall                                & \multicolumn{1}{c|}{0.91±0.03}                & \multicolumn{1}{c|}{0.84±0.10}  & \multicolumn{1}{c|}{0.73±0.06} & {\ul \textbf{0.93±0.03}} & \multicolumn{1}{l|}{0.66±0.0} & 0.65±0.02 & 0.98±0.02                       \\ \hline
\end{tabular}
\end{table*}

\subsubsection{\textbf{RQ1: How does CLog compare against baselines on the task of failure detection?}} 
We evaluate CLog detection performance against three unsupervised methods, two commonly used developer practices and one supervised method. The training is done on 60\% randomly sampled normal sequences, while the thresholds (and other hyperparameters) are selected on a random sample of 20\% normal sequences. The rest of the sequences are used to report the performance scores. The best results for CLog are obtained for a total of 10 subprocesses (centroids), two hidden states in the HMM and a window size of 180 seconds\footnote{The code and the data are given in the GitHub repository of the project \href{https://github.com/context-aware-Failure-Identification/CLog.git}{https://github.com/context-aware-Failure-Identification/CLog.git}}. The experiments are repeated ten times to reduce the assessment bias of the results. We report the mean and standard deviation of the results.

Since CLog is an unsupervised method, we first discuss the results between CLog and the unsupervised baselines. \tablename~\ref{tab:FD} shows the results. CLog outperforms the unsupervised baselines by margins between 9-17\% on the F1 score. Importantly, CLog and HMM both use HMM to model the sequences, but they differ in the granularity of the input representation. Marginalizing over the learning method suggest that changing the input representation of the log event sequences with sequences of subprocesses is beneficial. Combining these results with our observation (see \figurename~\ref{fig:motivation}) demonstrates that reducing the entropy by changing the input representation improves the detection performance. CLog predominantly improves the precision over the sequence-based methods (DeepLog and HMM), while having strong performance on recall. The input of the sequential-based baselines has larger entropy which challenges the discrimination against normal sequences, leading to many of them being detected as failures, i.e., increasing the false positives. Comparing CLog against the quantitative-based method (PCA) leads to good performance in precision but reduced recall. The count vector representation and the limited modeling power of PCA (as a linear model) are potential causes for the incorrect detection of the true failures. Notably, from an economic perspective, the observed improvements are significant because improving failure detection by even 0.1\% in F1 score can save hundreds of thousands of dollars~\cite{UniLog2021}.

The improved performance of CLog against the two commonly used developer practices, i.e., log level (by 24\% on F1) and single line semantic-based approach, is mainly due to the problem of insufficient logging failure coverage. The semantic-based baseline is constructed based on the expert inspection of the logs for failures, i.e., it detects all of the single-line logged failures. However, as we observed when performing the manual log analysis, and as shown in Cotroneto et al.~\cite{cotroneo2019bad}, around 20\% of the failures in the dataset are not explicitly logged. In comparison, CLog can detect non-logged failures because it models different contexts, i.e., its correlates events co-occurring together. The violations of these contexts (e.g., an expected log event is missing from the context) are informative in implicitly detecting non-logged failures. The log level-based approach experiences the lowest performance. Despite the problem of insufficient failure coverage, it further suffers from the problem of wrong log level assignment~\cite{Hassan2017}. A log may be assigned a log level "ERROR" but still describe a normal event. Therefore, relying on the log level leads to reporting more failures than there are, affecting the precision. 

Finally, comparing CLog against the supervised baseline LogRobust, suggests that CLog has a drop in performance by 2\% on the F1 score while for others, it exceeds 11\%. However, LogRobust requires labeled log sequences to build a model. Due to a large number of logs constantly being generated, the labeling is often infeasible in practice, and it is the most common referenced critique of the supervised methods~\cite{surveyLogAnalysis}. Therefore, CLog has better practical properties because of the high detection performance and unsupervised design. 

\begin{table}[!h]
\caption{Comparison of CLog against baselines on FTI.}
\label{tab:resFTI}
\resizebox{1\columnwidth}{!}{%
\begin{tabular}{c|c|cccc}
\hline
\multirow{2}{*}{Scores}    & \multirow{2}{*}{\begin{tabular}[c]{@{}c@{}}Multiclass\\ method\end{tabular}} & \multicolumn{1}{c|}{\begin{tabular}[c]{@{}c@{}}CLog\\ (pHMM rep.)\end{tabular}} & \multicolumn{1}{c|}{\begin{tabular}[c]{@{}c@{}}CLog\\ (CV rep.)\end{tabular}} & \multicolumn{1}{c|}{\begin{tabular}[c]{@{}c@{}}CLog\\ (CV+pHMM \\combined)\end{tabular}} & \begin{tabular}[c]{@{}c@{}}LogClass\\ (TFILF)\end{tabular} \\ \hline
\multirow{4}{*}{F1}        & RF                                                                           & \multicolumn{1}{c|}{0.74±0.0}                                                  & \multicolumn{1}{c|}{\textit{0.86±0.01}}                                          & \multicolumn{1}{c|}{\textit{0.86±0.01}}                                        & 0.84±0.05                                                  \\  
                           & DT                                                                           & \multicolumn{1}{c|}{0.74±0.0}                                                  & \multicolumn{1}{c|}{0.78±0.04}                                                   & \multicolumn{1}{c|}{0.78±0.03}                                                 & \textit{0.84±0.08}                                         \\  
                           & LR                                                                           & \multicolumn{1}{c|}{0.72±0.0}                                                  & \multicolumn{1}{c|}{0.86±0.0}                                                    & \multicolumn{1}{c|}{{\ul \textbf{0.87±0.0}}}                                   & 0.8±0.1                                                    \\  
                           & AdaBoost                                                                     & \multicolumn{1}{c|}{0.69±0.0}                                                  & \multicolumn{1}{c|}{0.86±0.02}                                                   & \multicolumn{1}{c|}{{\ul \textbf{0.87±0.02}}}                                  & 0.62±0.12                                                  \\ \hline
\multirow{4}{*}{Precision} & RF                                                                           & \multicolumn{1}{c|}{0.74±0.0}                                                  & \multicolumn{1}{c|}{\textit{0.86±0.01}}                                          & \multicolumn{1}{c|}{\textit{0.86±0.01}}                                        & 0.83±0.05                                                  \\  
                           & DT                                                                           & \multicolumn{1}{c|}{0.74±0.0}                                                  & \multicolumn{1}{c|}{0.78±0.04}                                                   & \multicolumn{1}{c|}{0.78±0.03}                                                 & \textit{0.82±0.07}                                         \\  
                           & LR                                                                           & \multicolumn{1}{c|}{0.71±0.0}                                                  & \multicolumn{1}{c|}{0.85±0.0}                                                    & \multicolumn{1}{c|}{{\ul \textbf{0.87±0.0}}}                                   & 0.77±0.08                                                  \\  
                           & AdaBoost                                                                     & \multicolumn{1}{c|}{0.71±0.0}                                                  & \multicolumn{1}{c|}{0.86±0.02}                                                   & \multicolumn{1}{c|}{0.86±0.02}                                                 & 0.59±0.13                                                  \\ \hline
\multirow{4}{*}{Recall}    & RF                                                                           & \multicolumn{1}{c|}{0.75±0.0}                                                  & \multicolumn{1}{c|}{0.87±0.01}                                                   & \multicolumn{1}{c|}{0.87±0.01}                                                 & \textit{0.86±0.06}                                         \\  
                           & DT                                                                           & \multicolumn{1}{c|}{0.75±0.0}                                                  & \multicolumn{1}{c|}{0.81±0.04}                                                   & \multicolumn{1}{c|}{0.81±0.03}                                                 & \textit{0.85±0.07}                                          \\  
                           & LR                                                                           & \multicolumn{1}{c|}{0.73±0.0}                                                  & \multicolumn{1}{c|}{0.88±0.0}                                                    & \multicolumn{1}{c|}{{\ul \textbf{0.89±0.0}}}                                   & 0.88±0.11                                                  \\  
                           & AdaBoost                                                                     & \multicolumn{1}{c|}{0.73±0.0}                                                  & \multicolumn{1}{c|}{0.87±0.02}                                                   & \multicolumn{1}{c|}{0.88±0.02}                                                 & 0.74±0.14                                                  \\ \hline
\end{tabular}%
}
\end{table}

\subsubsection{\textbf{RQ2: How effective is CLog for the problem of failure type identification?}}
This RQ evaluates the capability of the FTI module of CLog to reuse the historical information from the operator in detecting different types of failures. Specifically, we evaluate three representations of the subprocess sequences for CLog (1. probability score from the HMM (pHMM), 2. count vectors (CV), and 3. combination of both) against LogClass~\cite{LogClass2021} as a baseline. LogClass uses single logs as input. We randomly sample 60\% of the labeled failure sequences/logs from the original dataset to train the multiclass model, while the remaining 40\% are used for evaluation. To reduce the bias due to the sampling, we repeated the experiments 30 times. We report the average performance scores and their standard deviation over the different methods, evaluating the representations independent of the methods. 

\tablename~\ref{tab:resFTI} enlists the results of the three different representations of CLog and the baseline on the FTI subproblem. The analysis of the three representations by CLog suggests that the combination (CV+pHMM) achieves the best F1 score. Predominantly, the improvement originates from the count vectors, seen by the better individual results in comparison with pHMM. Finally, the combination (CV+pHMM) of CLog outperforms the baseline LogClass. LogClass uses single logs to identify the type of failures. Therefore, if the failure is not explicitly logged, LogClass cannot identify its type. On the contrary, CLog considers the occurrence of the individual subprocesses and can represent discriminative patterns among the types of failures improving the performance. 

\begin{figure}[!t]
\centering
\includegraphics[width=0.45\textwidth]{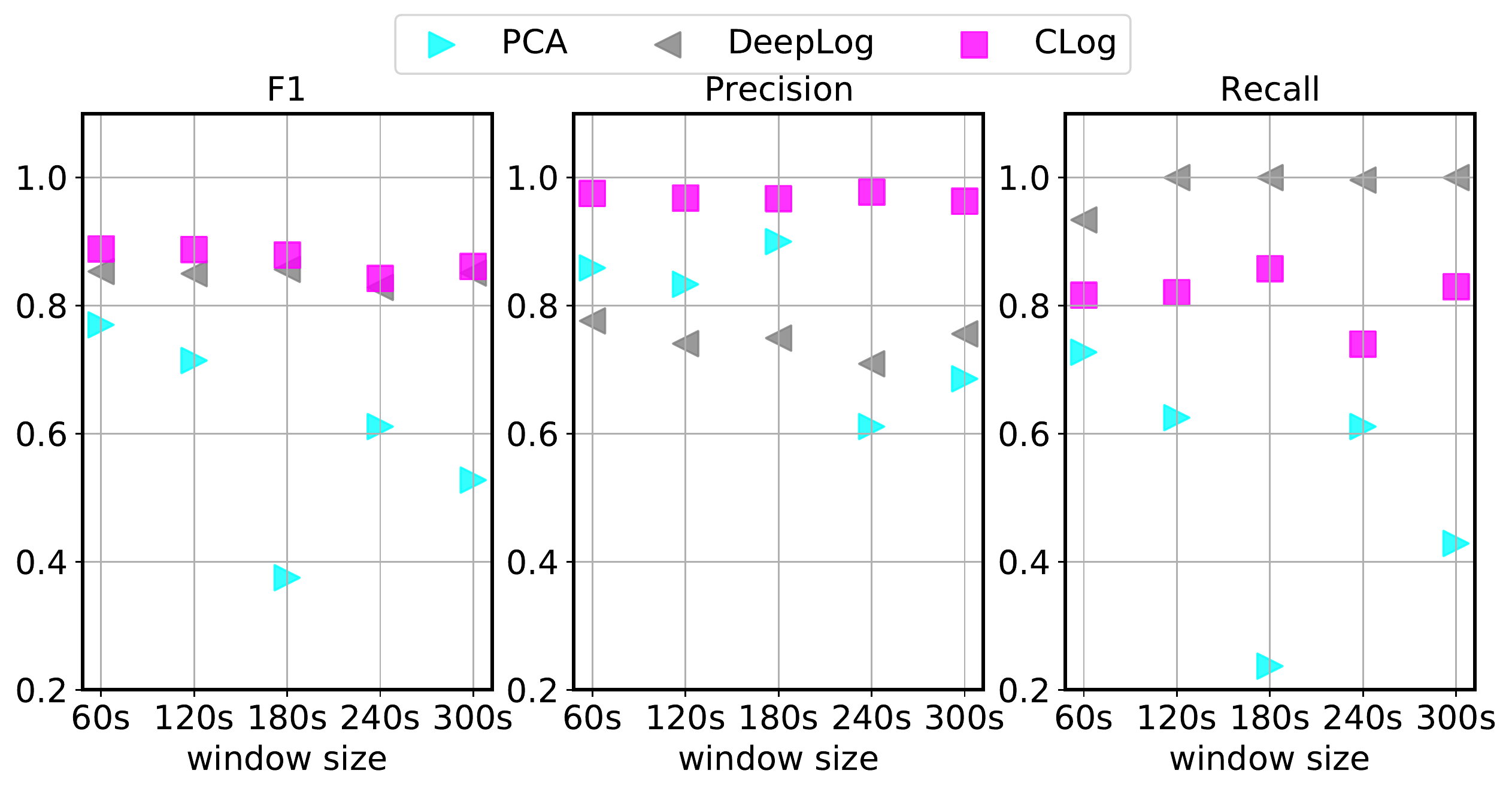}
\caption{Impact of the window size over the detection performance.}
\label{fig:resultsFD}
\end{figure}

\subsubsection{\textbf{RQ3: How does the entropy of the sequences influence the failure detection performance independent of the model?}}
With this question, we verify the impact of the unstable log event sequences over the failure detection performance in a model agnostic manner. Following our key observation given in Section~\ref{keyObservation}, we grouped the input events into time-intervals of increasing window size and evaluated several models of CLog, DeepLog, and PCA similarly as in RQ1. The best average performance of CLog averaged over the window sizes (not individual as in RQ1) is obtained for 30 subprocesses.

\figurename~\ref{fig:resultsFD} shows the results. It can be observed that as the window size increases, the detection performance decreases. Paring the average entropy over the sequences for different window sizes (see \figurename~\ref{fig:motivation}) with the detection results reveals a negative correlation between the increased entropy and failure detection performance. Owning to the greater modeling power CLog, and DeepLog show a relatively lower drop in performance in the range 4-5\%. In comparison, PCA has a performance drop between 8-50\%. Importantly, CLog outperforms DeepLog because of the smaller instability in the input.

\subsubsection{\textbf{RQ4: How robust is CLog on the problem of failure detection?}}
We conduct experiments on the synthetic dataset. The failure detector was trained on the original data as in RQ1. We randomly inject $b-$percentages unstable sequences using the synthetic dataset generation procedure previously described. \tablename~\ref{synteticData} shows the results. CLog's detection method preservers high detection performance even under a high ratio of unstable sequences. The results demonstrate that the extracted subprocesses are sufficiently sensitive to the local changes within the log sequences, making the performance robust.
\begin{table}[!h]
\centering
\caption{CLog Failure Detection Evaluation on Synthetic Data}
\label{synteticData}
\begin{tabular}{c|c|c|c}
\hline
injection ratio & F1   & Precision & Recall \\ \hline
5\%  & 0.94 & 0.97      & 0.91   \\ 
10\% & 0.92 & 0.95      & 0.89   \\ 
15\% & 0.90 & 0.95      & 0.86   \\ 
20\% & 0.88 & 0.94      & 0.83    \\ \hline
\end{tabular}
\end{table}

\section{Related Work}\label{rw}
\textbf{Failure Detection.} There are plenty of works considering the problem of log-based failure detection. Considering the assumption of whether log labels are available or not, the methods are categorized into: unsupervised and supervised. We first discuss the unsupervised methods. They are considered practically useful because they do not require labels~\cite{surveyLogAnalysis}. Predominantly, these are one-class methods modeling the normal system state and reporting failures when significant deviations occur. Yamanishi et al.~\cite{KDD2005} introduce an unsupervised sequential method for failure detection that uses HMM on log event sequences to model the normal state. The probability under the HMM is used as a normality score. DeepLog~\cite{DeepLog} trains a neural network -- LSTM on sequences of log events on the auxiliary task of \textit{next event prediction}. If the output prediction of the auxiliary task is wrong, the method reports failure. LogAnomaly~\cite{LogAnomaly} is similar to DeepLog, but further augments the input of DeepLog with semantic and count-based features. Another popular method is PCA~\cite{PCA}. It is a reconstruction-based method that constructs subspace from the count vectors of the normal log event sequences and uses the reconstruction error to detect failures. Compared to other methods that do not directly address the problem of unstable sequences, CLog addresses it by representing the sequences of log events with sequences of subprocesses. The semantic-based methods use the semantics of the single logs to identify failures. Despite the traditionally used approaches in industry, like keyword search (e.g., search for "error") or log level search~\cite{LogCluster2016}, another line of works learns properties of failure logs. One such example is Logsy~\cite{Logsy}. It combines labeled data from other software systems and a hyperspherical loss when learning the discriminative properties of the failures. The semantic baseline we considered assumes that methods like Logsy perform ideally, thereby, we do not directly compare with it. These methods are unable to detect failures that not explicitly are logged. By modeling sequences, CLog detects contextual failures when abrupt changes in the normality contexts of the co-occurring events occur.

Supervised methods assume the availability of labels for logs from the target system. LogRobust~\cite{deepLogAnomalySurvey} uses a sequential representation of the input to train a deep learning neural network -- LSTM augmented with an attention mechanism to learn failure sequences. Other methods, such as SVM, decision trees, logistic regression and nearest neighbours, are also being considered~\cite{Branisova2015}.~Due to the evolution of logs and their large volumes, the expensiveness of labeling is referenced critique to supervised methods, questioning their usability~\cite{surveyLogAnalysis}. 

\textbf{Failure Type Identification.} In one of the earliest works, Oliner et al.~\cite{SuperComputer} use keyword search of common words (e.g., "interface failure", "error", and similar) to identify different types of failures within single logs of four different supercomputer systems. Similarly, Meng et al.~\cite{Meng2018} use single log lines represented as bag-of-words, alongside the Random Forest method to classify different types of system logs. LogClass~\cite{LogClass2021} introduces TF-ILF as a novel representation method for individual logs and applies commonly used multiclass classification methods to categorize the type of failure. A common drawback of these approaches is the assumption of full failure coverage in single logs. However, due to the problem of insufficient logging coverage within the source code~\cite{He2020SurveyLogMining}, this may not always be the case.  Different from others, we pair count vectors from the subprocesses of a given sequence with a multiclass classifier to use the past information about similar failure types.

\section{Conclusion}\label{conc}
This paper addresses the problem of the automation of log-based failure identification, which is a crucial maintenance task to enhance the reliability in cloud systems. It introduces a novel method CLog, which decouples the problem of failure identification into two subproblems 1) failure detection and 2) failure type identification. We observe that by representing the input log data as sequences of subprocesses instead of sequences of individual events, the entropy in the input, caused by the unstable logs, is reduced. CLog uses this observation and introduces a novel subprocess extraction method, which jointly trains context-aware deep learning and clustering methods to extract subprocesses. Our experiments demonstrate that the extracted sequences of subprocesses are beneficial for improving the performance of the two subproblems of 1) failure detection (by 9-24\% over the baselines) and 2) failure type identification (by 7\% over the baseline). Further, we show that CLog has robust performance, under high ratios of injected unstable sequences, experiencing just a 6\% performance drop. The key observation presented herein opens new possibilities for how to most efficiently extract meaningful subprocess with minimal information about the sequences (e.g., discarding the sequence identifiers), which we aim to explore next. These achievements can ultimately bridge the gap between automatic log-based failure detection and root-cause analysis, further enhancing the reliability of cloud systems. 

\bibliographystyle{IEEEtran}
\bibliography{IEEEabrv,referencesValid.bib}

\end{document}